\documentstyle[multicol,aps]{revtex}
\begin{document}
\draft
\title{Ideal Spin Filters: A Theoretical Study of Electron  Transmission
Through Ordered and Disordered Interfaces Between Ferromagnetic Metals
and Semiconductors}
\author{George Kirczenow}
\address{Department of Physics, Simon Fraser University, Burnaby, B.C.,
Canada  V5A 1S6}
\maketitle
\begin{abstract}

It is predicted that certain atomically ordered interfaces between some
ferromagnetic metals (F) and semiconductors (S) should act as ideal spin
filters that transmit electrons only from the majority spin bands or only
from the minority spin bands of the F to the S (and from the S only to
the  majority spin bands or only to the  minority spin bands of the F) at
the Fermi energy, even for F with both majority and minority bands at the
Fermi level. Criteria for determining which combinations of F, S and
interface should be ideal spin filters are formulated. The criteria
depend only on the bulk band structures of the S and F and on the
translational symmetries of the S, F and interface. Several examples of
systems that meet these criteria to a high degree of precision are
identified. Disordered interfaces between F and S are also studied and it
is found that intermixing between the S and F can result in interfaces
with spin anti-filtering properties, the transmitted electrons being much
{\em less} spin polarized than those in the ferromagnetic metal at the
Fermi energy. 

\end{abstract}
\begin{multicols}{2}

\section{Introduction}

In ferromagnetic condensed matter systems
the spin up and spin down states are occupied asymmetrically by
electrons. Because of this asymmetry it is possible for an
applied electric field to drive a spin-polarized electron current
across the interface between a ferromagnet and a non-magnetic
material. Spin-polarized electron transport has been achieved
experimentally from ferromagnetic metals to superconductors by
Meservey, Tedrow and Fulde\cite{Mes}, from ferromagnetic to
normal metals by Johnson and Silsbee\cite{J&S},  between
ferromagnetic metals separated by thin insulating 
films by Julliere\cite{Julliere}, and
from magnetic semiconductors to non-magnetic 
semiconductors by Fiederling {\em et al.}\cite{Fiederling} and
Ohno {\em et al.}\cite{Ohno}. 

Injection of strongly spin-polarized
electron currents from ferromagnetic metals into semiconductors
has also long been recognized as an important fundamental goal in
condensed matter  physics\cite{long}. Attaining it would 
have a significant technological impact in the area of
spintronics, the branch of electronics that  utilizes the
electron's spin  degree of freedom as well as its charge  to
store, process and transmit information.  However only weak
signatures of
spin-polarization of electrons injected from ferromagnetic metals
into semiconductors through the metal-semiconductor interface
have been reported \cite{Lee,Hammar} 
and the interpretation of such
experiments is controversial
\cite{Monzon,Wees,Hammar2}. 

By extending previous theoretical
work    by van Son, van Kempen and Wyder\cite{vS} and by Valet
and Fert
\cite{VF} on spin transport in metallic systems, Schmidt {\em et
al.}\cite{Schmidt} recently concluded that for devices in the
diffusive transport regime only a weak ($<0.1\%$) 
spin-polarization of electrons injected  from a ferromagnetic
metal into a semiconductor is possible, even  in principle, unless
the ferromagnetic contact is almost 100\% spin-polarized, which
is not the case for such common ferromagnetic metals such as
Fe, Co, Ni and permalloy. The essential reason was that 
in an electric circuit consisting 
of a diffusive semiconductor in series with a metal, 
the net resistance of the
circuit is dominated by the resistance 
of the semiconductor which is spin-independent, 
and therefore the 
spin up and down currents flowing through the semiconductor
should be almost equal. 
However, Schmidt {\em et
al.}\cite{Schmidt} did not allow 
in their analysis for the possibility that
electron transmission through the interface between the 
ferromagnet and semiconductor may be very strongly spin-dependent.

In this article the spin-dependence
of electron transmission from ferromagnetic metals to semiconductors
is examined theoretically. It is predicted that, in contrast to the 
interfaces
between ferromagnetic and normal metals and to the tunnel barriers
between ferromagnetic metals previously discussed in the
literature\cite{Bruno,Schep,Stiles,Tsymbal,Schep2,MacLaren},  
atomically ordered and suitably
oriented  interfaces between some ferromagnetic metals and some 
semiconductors should be {\em almost perfect spin filters.} 
That is, they should
transmit only majority or only minority spin band electrons from the
ferromagnet to the semiconductor even for ferromagnetic 
metals for which both majority and minority spin bands
are present at the Fermi level. Such spin filters
make it possible, in principle, to overcome the difficulties discussed
by Schmidt {\em et al.}\cite{Schmidt} and to achieve injection of
strongly spin-polarized electric currents from ferromagnetic metals
into semiconductors. Several examples of combinations of
ferromagnetic metals, semiconductors and interfaces that are good 
candidates 
for near-ideal spin filters are
identified in this article. 

Recently it has been proposed by Ferreira  {\em et al.} that a
superlattice that consists of alternating layers of two different
materials arranged in a periodic sequence may be a perfect spin filter
for electron transmission between two ferromagnets\cite{superlat} and
that a pair of magnetic superlattices connected by a conducting medium
with  a low carrier density may also be a perfect spin
filter\cite{superlat2}. These possibilities were explored using simple
free electron models and Kronig-Penney-like
potentials\cite{superlat,superlat2}.  By contrast the ideal spin filters
introduced here require only a {\em single} interface for their operation
and the spin-polarized electron transmission is between a ferromagnet and
semiconductor.  Also the present work takes into account the atomic
crystal structures and realistic electronic band structures of the
materials involved.

The present approach to injecting spin-polarized  electrons from
ferromagnetic metals into semiconductors also differs fundamentally from
a previous suggestion  by Johnson
\cite{Johnson,Hammar2} that was based on the Rashba effect in 
quasi-two-dimensional electron gases.

The effects of disorder at the interface on the spin polarization of
electrons transmitted from ferromagnetic metals to semiconductors are
also addressed in the present work by solving a simple  tight binding
model numerically. The model exhibits perfect spin filtering for an
ordered interface between the semiconductor and ferromagnetic metal.
Intermixing between the semiconductor and metal at the interface
drastically reduces the spin-polarization of the transmitted electron
flux. When intermixing completely destroys the interface symmetries that
result in spin filtering, the spin-polarization of the transmitted
electron flux does not resemble that of the ferromagnetic metal at the
Fermi level; it is very much weaker.  This effect has a different physical
origin from that discussed by Schmidt {\em et al.}\cite{Schmidt}.
  
The article is organized as follows: In Section \ref{Rules} general
selection rules for electron transmission through atomically ordered
interfaces between crystals are derived in a form suited  to the present
purpose. Criteria for identifying possible candidates for ideal spin
filters are then formulated based on these selection rules. 

The selection rules derived in Section \ref{Rules} are exact and apply 
not only to interfaces between ferromagnetic metals and semiconductors
(the subject of the present work) but also to ordered interfaces
between crystalline materials in general. Theoretical work on electron
transmission through the ordered interfaces between various crystalline 
materials, based on {\em ab initio} computer calculations and 
analytic models has been published by several authors \cite
{Bruno,Schep,Stiles,Tsymbal,Schep2,MacLaren,superlat,superlat2,StilesandHamann,StilesandPenn}.
The general form of the selection rules that is derived in Section
\ref{Rules} and is needed for the present purpose does not appear in
those publications but is consistent with the formalism
underlying the {\em ab initio} calculations.         

In Section
\ref{examples} some examples of combinations of  ferromagnetic metals,
semiconductors and interfaces that are candidates for nearly ideal spin
filters are identified and discussed.  The effects of disorder at the
interface on the spin polarization of the transmitted current are
examined in Section \ref{Disorder}. A summary and some further comments
are contained  in Section \ref{Conclusions}.

\section{Selection rules and criteria for ideal spin filters}
\label{Rules}      

When an electron is transmitted through an ordered interface
between two crystals, the projection of its Bloch state 
wave vector onto the interface is conserved
up to reciprocal lattice vectors. In this section I derive a 
precise formulation of this principle that applies to
electron transmission through
general semiconductor-ferromagnetic metal interfaces. Based on
this formulation, I then define criteria for identifying 
combinations of ferromagnetic metals, semiconductors and
interfaces that are candidates for ideal spin filters.    

Consider an atomically ordered plane interface between two
crystals, a semiconductor $S$ and a ferromagnetic metal 
$M$. The two crystals
may be in direct contact with each other at the interface 
or the interface may include one or more ordered layers 
of other atomic species than those present in $S$ and $M$
and/or the same species in a different spatial arrangement. 

Far from the interface the periodicities of the two crystals 
are described by their three-dimensional sets of Bravais lattice 
vectors \{${\bf R}_S$\} and \{${\bf R}_M$\}, respectively. The
periodicity of the entire system consisting of the two
crystals and the interface is described by a
two-dimensional Bravais lattice of symmetry translations  
\{${\bf R}_I$\} parallel to the plane of
the interface. The corresponding three and two-dimensional 
reciprocal lattices are the sets of vectors 
\{${\bf K}_S$\}, \{${\bf K}_M$\}
and \{${\bf K}_I$\}, respectively. 

Because of the symmetry of the entire system under the 
set of translations 
\{${\bf R}_I$\} parallel to the plane of
the interface, a
complete set of one-electron energy eigenstates of the entire system
can be chosen in the Bloch form
\begin{equation}
 {\Psi}_{{\bf k}_I s} ({\bf r}) = 
e^{i{\bf k}_I \cdot{\bf r}}u_{{\bf k}_I s}
({\bf r}) 
\label{eq:Bloch}
\end{equation}
where ${\bf r}$ is the position of the electron, ${\bf k}_I$ 
is a vector parallel
to the plane of the interface and $u$ can be written in the form $
u_{{\bf k}_I s} ({\bf r})= \sum_{{\bf K}_I} 
\Lambda^{{\bf k}_I s}_{{\bf K}_I}
({\bf r}_\perp)e^{i{{\bf K}_I}\cdot{\bf r}}$.  
Here the Fourier coefficients $\Lambda^{{\bf k}_I s}_{{\bf K}_I}$ 
depend on ${\bf
r}_\perp$, the component of ${\bf r}$ in the direction orthogonal 
to the plane
of the interface. The states
${\Psi}_{{\bf k}_I s}$ include among them the scattering states 
of electrons that 
are incident on the interface from the ferromagnetic metal
crystal at the Fermi energy and are partly 
or completely transmitted and/or reflected at the interface.
Deep in the ferromagnet these scattering states can be written as linear 
combinations of the Bloch states ${\psi}_{{\bf k}_M
\tilde{s}}$ of the (three-dimensional) ferromagnetic metal crystal at the
Fermi energy. I.e.,
deep in the ferromagnet
\begin{equation}
 {\Psi}_{{\bf k}_I s} ({\bf r}) = \sum_{{\bf k}_M,\tilde{s}} 
A^{{\bf k}_I s}_{{\bf k}_M \tilde{s}}
{\psi}_{{\bf k}_M \tilde{s}} ({\bf r}) 
\label{eq:Blochrel}
\end{equation}
Writing the Bloch states ${\psi}_{{\bf k}_M
\tilde{s}}$ of the ferromagnet
 in the Fourier form ${\psi}_{{\bf k}_M \tilde{s}}({\bf r})=
\sum_{{\bf K}_M} 
\lambda^{{\bf k}_M \tilde{s}}_{{\bf K}_M}e^{i({{\bf k}_M}+{{\bf
K}_M})\cdot{\bf r}}$
and combining (\ref{eq:Bloch}) with (\ref{eq:Blochrel}) then yields
\begin{equation}
 \sum_{{\bf K}_I} 
\Lambda^{{\bf k}_I s}_{{\bf K}_I}({\bf r}_\perp)e^{i({{\bf k}_I}+{{\bf
K}_I})\cdot{\bf r}}
= \sum_{{\bf k}_M,{\bf K}_M,\tilde{s}} 
A^{{\bf k}_I s}_{{\bf k}_M \tilde{s}}
\lambda^{{\bf k}_M \tilde{s}}_{{\bf K}_M}e^{i({{\bf k}_M}+{{\bf
K}_M})\cdot{\bf r}}  
\label{eq:Blochequality}
\end{equation}
for ${\bf r}$ deep in the ferromagnet. 
Equation (\ref{eq:Blochequality}) can only be satisfied 
for all ${\bf r}$ deep in the ferromagnet if for
some ${\bf k}_M$ on the ferromagnet's Fermi surface and for some 
reciprocal lattice vectors ${{\bf K}_I}$ and ${{\bf K}_M}$
\begin{equation}
{\bf k}_I = ({\bf k}_M + {{\bf K}_M})_\| - {{\bf K}_I}
\label{eq:intmetcond}
\end{equation}
where $(\cdots)_\|$ denotes projection onto the plane of the interface.
Similarly, deep in the semiconductor the same scattering states
can be expressed in terms of semiconductor Bloch states yielding 
instead of
(\ref{eq:intmetcond})   
\begin{equation}
{\bf k}_I = ({\bf k}_S + {{\bf K}_S})_\| - {{\bf K'}_I}
\label{eq:intsem}
\end{equation}
where ${\bf k}_S$ is a vector on the Fermi surface of the semiconductor
and ${{\bf K'}_I}$ is a vector of the two-dimensional 
reciprocal lattice of the
entire system. For transmission from the ferromagnet to the 
semiconductor to
occur, both (\ref{eq:intmetcond}) and (\ref{eq:intsem}) must be
satisfied and thus
\begin{equation}
({\bf k}_S)_\| = ({\bf k}_M)_\| + ({{\bf K}_M} - {{\bf K}_S})_\| -
{{\bf K}_I}+ {{\bf K'}_I}.
\label{eq:metsem}
\end{equation}

Let the word `projection' stand for
`projection onto the plane of the interface'. Then equation 
(\ref{eq:metsem}) 
implies the following selection rule: Transmission  of electrons at the
Fermi energy is forbidden from the  majority (minority) spin bands of
the ferromagnet to the semiconductor (and vice-versa) unless the
projections of the Fermi surfaces of the semiconductor and of the
majority (minority) spin bands of the ferromagnetic metal are
connected by a vector that is the sum of a (2D) reciprocal lattice
vector of the entire system  and projections of reciprocal lattice
vectors of the semiconductor and ferromagnet.

It should be noted that the above selection rule depends only on
the  bulk electronic structure of the ferromagnetic metal and
semiconductor and on translational symmetries, and not 
on the details of the electronic structure of the interface. 
 
If in this way transmission of majority spin band 
electrons from the ferromagnet to the
semiconductor is allowed but that of minority 
spin band electrons is forbidden (or vice-versa)
then, in the absence of spin-flip scattering and if spin-orbit
coupling can be neglected (see Section \ref{Conclusions}), 
the system is an ideal spin
filter for injection of spin-polarized electrons from the ferromagnet
into the semiconductor. 

While the above derivation of the criteria for ideal spin filters 
also applies to interfaces between normal and ferromagnetic metals,
the Fermi surfaces of most metals enclose large enough fractions of the 
Brillouin zone that the criteria cannot be satisfied. 
On the other hand the Fermi surface of a
semiconductor encloses  only a very small fraction 
of the Brillouin zone. Because of this
some combinations of ferromagnetic metal, semiconductor and interface
are possible candidates for nearly ideal spin filters as will be
seen below.

\section{Some candidates for nearly ideal spin filters}
\label{examples}

\subsection{The simplest case}
\label{simplicity}

The simplest semiconductors to consider in the present context are 
those with a single lowest conduction band minimum 
(for n-type materials) or
highest valence band maximum (for p-type materials\cite{p-type})
located at the center of the Brillouin zone (i.e. at ${\bf k}=0$) so that
$({\bf k}_S)_\| =0$ in equation (\ref{eq:metsem}). 

For such semiconductors it is helpful 
to examine separately the case where $({{\bf K}_M})_\| = ({{\bf K}_S})_\| 
= {{\bf K}_I} = {{\bf K'}_I} = 0$ in equation (\ref{eq:metsem}) 
so that equation (\ref{eq:metsem}) 
reduces to 
\begin{equation}
({\bf k}_M)_\| = 0
\label{eq:metsem0}
\end{equation}
and the complimentary case where one or more of $ ({{\bf K}_M})_\|, ({{\bf K}_S})_\|, 
{{\bf K}_I}$ and ${{\bf K'}_I}$ is not zero so that
equation (\ref{eq:metsem}) becomes
\begin{equation}
({\bf k}_M)_\| = {{\bf K}_I} - {{\bf K'}_I} - ({{\bf K}_M} - {{\bf K}_S})_\|.
\label{eq:metsemnot0}
\end{equation}

The criteria for ideal spin filters derived in Section \ref{Rules} then reduce to
the following:\\
(1) Neither equation (\ref{eq:metsem0}) nor equation (\ref{eq:metsemnot0}) should be satisfied
(for any choice of $ {{\bf K}_M}, {{\bf K}_S}, 
{{\bf K}_I}$ and ${{\bf K'}_I}$) for any ${\bf k}_M$ on the majority (minority) spin Fermi 
surface of the metal\\
{\em and}\\
(2) Either equation (\ref{eq:metsem0}) or equation
(\ref{eq:metsemnot0})  (for some choice of $ {{\bf K}_M}, {{\bf K}_S}, 
{{\bf K}_I}$ and ${{\bf K'}_I}$) or
both should be satisfied for some ${\bf k}_M$ on the minority (majority) spin Fermi 
surface of the metal.

Whether equation (\ref{eq:metsem0}) is satisfied depends only on the Fermi surface
geometry of the metal and the orientation of the interface but {\em not} on the crystal
structure of the semiconductor, metal or interface. This greatly simplifies the process of
screening for systems involving direct gap semiconductors that may be nearly ideal
spin filters: Start by identifying as possible candidates those combinations of
ferromagnetic metal and interface orientation for which equation (\ref{eq:metsem0}) is {\em
not} satisfied for the majority spin Fermi surface, the minority spin Fermi surface or both.
Having narrowed the field of potential candidates in this way, proceed with detailed
analyses of the crystal structures of specific combinations of materials at suitably
oriented interfaces to determine whether the remaining conditions for ideal spin filters that
involve equation (\ref{eq:metsemnot0}) and equation (\ref{eq:metsem0}) are also
satisfied.        

Inspection of the calculated
band structures and Fermi surfaces of some common ferromagnetic
metals that are available in the literature
\cite{Papaconst,Mertig,web} 
shows that equation (\ref{eq:metsem0}) is not satisfied for the majority spin
Fermi surface of hcp Co if the interface is 
orthogonal to the (001)
crystallographic axis, i.e., parallel to a basal plane of 
hexagonally close packed Co atoms. This is also the case for fcc Ni and fcc Co 
for interfaces perpendicular to their (111)
crystallographic axes\cite{exception}. Some other ferromagnetic metals
whose published electronic band structures 
\cite{web,Sanchosuperl,Zhao,Sundararajan,Prins,Ahuja,Schepsuperl} also 
do not satisfy equation (\ref{eq:metsem0}) for majority and/or minority spin
electrons and some orientation(s) of a
putative interface plane include 
simple cubic Mn, CoS$_2$, FeAl, $\tau$-MnAl, Gd, Tb and some magnetic
superlattices.

While consideration of equation (\ref{eq:metsem0}) can be a useful starting point, 
a more detailed analysis of
specific systems consisting of the ferromagnetic metal, semiconductor
and interface is essential to determine whether they satisfy {\em all} of the
criteria for ideal spin-filters derived in Section \ref{Rules}. 
Such analyses will be
outlined below for  a number of systems involving 
the ferromagnetic metals Co, Ni, 
CoS$_2$, FeAl, $\tau$-MnAl, Gd, Tb, Pd$_3$Fe, Co$_3$Pt 
and some magnetic superlattices
together  with a variety of n- and p-type direct and 
indirect gap  semiconductors.

\subsection{Spin filters involving the ferromagnetic metals 
hcp Co, fcc Ni or fcc Co}
\label{CoNi}

\subsubsection{Semiconductors with Fermi surfaces at the center of the 
first Brillouin zone}

The (001) crystallographic planes of hcp Co and the (111) planes of fcc
Ni and fcc Co  consist of metal atoms in a hexagonal close-packed
arrangement. The (111) atomic planes of semiconductors with the diamond
and zinc blende crystal structure also have hexagonal atomic
arrangements, but in most cases with considerably larger in-plane nearest
neighbor atomic spacings than those of the metals.  Direct gap
semiconductors with the wurtzite structure also have planes of atoms in a
hexagonal arrangement, and the in-plane nearest neighbor spacings are
again in most cases significantly larger than those of the metals.
However, for many of these semiconductors, these in-plane nearest-neighbor
atomic spacings  are larger than those in the metals by factors close to
$\sqrt{3}$. For such semiconductors approximate atomic registry between
the hexagonal atomic planes of the metal and semiconductor can be
achieved by a rotation of the metal Bravais lattice relative to that of
the semiconductor through a $30^\circ$ angle about the axis perpendicular
to the plane of the interface.

The reciprocal lattice vectors and projections of reciprocal lattice
vectors onto the plane of the interface that enter equation
(\ref{eq:metsem}) are shown schematically in Fig.\ref{fig01} for
interfaces between hcp Co, fcc Ni and fcc Co, and semiconductors with the
diamond, zinc blende and wurtzite structures that are perfectly lattice
matched to the metals as described in the preceding paragraph. The
projections onto the plane of the interface of the reciprocal lattice
vectors of the metals are represented by the filled circles. The
projections onto the plane of the interface of the reciprocal lattice
vectors of the semiconductors matched to the metals are indicated by both
the open and filled circles. The open and filled circles also indicate
the reciprocal lattice vectors that are  associated with the group of
symmetry translations (parallel to the interface plane) of the whole
system consisting of the two crystals and the interface between them,
assuming that no lattice reconstruction occurs at the interface. The
hexagon (shown for reference) is the boundary of the projection of the
first Brillouin zone of  hcp Co onto the interface plane. For a direct
gap semiconductor whose conduction band minimum (valence band maximum) is
located at the center of the Brillouin zone, the open and filled circles
also  indicate the locations of the replicas of the conduction band
minimum (valence band maximum) in the repeated zone scheme. Inspection of
the calculated band structures and Fermi surfaces of hcp Co, fcc Ni, and
fcc Co\cite{Papaconst,Mertig,web}  shows that the projection of the Fermi
surface of the majority spin electrons in each of these metals onto the
interface plane does not overlap any of these replicas of the
semiconductor  conduction band minimum (valence band maximum), whereas
this is not  true of the Fermi surfaces of the minority spin electrons.
Therefore according to the reasoning in Section
\ref{Rules} only minority spin band electrons can be transmitted through
the interface at the Fermi energy and the interface is an ideal spin
filter if the semiconductor is perfectly lattice matched to the metal in
the above sense\cite{exception}. (Note that consideration of equation 
(\ref{eq:metsem0}) is by itself {\em not} sufficient to determine
whether these systems should be ideal
spin filters since equation 
(\ref{eq:metsem0}) does not address whether the projections of the
majority or minority Fermi surfaces of the metals onto the interface
plane overlap the projections of the semiconductor reciprocal lattice
vectors marked by open circles in Fig.\ref{fig01}; in an analysis
based on the form of the theory described 
in Section \ref{simplicity}, equation 
(\ref{eq:metsemnot0}) must be considered as well.)     

Some direct gap semiconductors with the zinc blende and wurtzite
structure that approximately lattice-match the metals in the above sense
(and therefore are candidates for spin filters when either n- or p-doped)
are listed below.  In each case the name of the semiconductor is followed
in parentheses by the ratio $a_{\rm p}/a$ (for hcp Co, fcc Ni and fcc Co,
respectively) of the value $a_{\rm p}$ of the lattice parameter required
for a perfect match with the metal, to the actual value $a$ of the
lattice parameter for the semiconductor.
 
Semiconductors with the zinc blende structure:\\ 
ZnTe (1.006, 1.000, 1.006),
GaSb (1.007, 1.001, 1.007),
InAs (1.014, 1.008, 1.013),\\
CdSe (1.015, 1.009, 1.014),
CuI  (1.016, 1.010, 1.016),
InP  (1.046, 1.040, 1.046),\\
InSb (0.948, 0.942, 0.947),
CdTe (0.947, 0.941, 0.946),
CdS  (1.054, 1.048, 1.054),\\
ZnSe (1.083, 1.077, 1.083),   
GaAs (1.086, 1.080, 1.086).
 
Semiconductors with the wurtzite structure:\\
CdSe (1.010, 1.004, 1.010),
CdS  (1.050, 1.044, 1.049).

An indirect gap semiconductor with the zinc blende structure 
whose valence band maximum is at the center of the Brillouin zone is
AlSb (1.001, 0.995, 1.000). As a p-type semiconductor it is also a
potential spin filter in conjunction with the the same metals.

For an interface to function as a nearly ideal spin filter, an accurate
lattice match is clearly desirable. In the above list the accuracy of the
lattice matching varies from excellent to marginal, depending on the
materials involved. It may be improved by alloying different
semiconductors, for example GaSb or InAs with InSb, at the expense of
introducing random defects. There is reason to expect that in at least
some cases such defects will not severely degrade the performance of spin
filters (which depends on the conservation of projected Bloch state wave
vectors up to reciprocal lattice vectors at the interface) since there
exist heterostructures such as GaAs/Al$_x$Ga$_{1-x}$As in which a 2D
electron gas can have very high mobilities despite being in contact with
such a semiconductor alloy. Another way to improve the lattice matching
is to grow very thin films (of the metal on the semiconductor or of the
semiconductor on the metal) in which the metal and semiconductor are in
perfect atomic registry with each other at the interface although the
thin film is elastically strained. 

\subsubsection{Hexagonal Boron Nitride}

The above examples have been of spin filters based on semiconductors
whose relevant conduction band minimum or valence band maximum is located
at the center of the Brillouin zone. However the selection rules and
criteria for ideal spin filters developed in Section \ref{Rules} also
apply to interfaces between ferromagnetic metals and semiconductors with
relevant band extrema that are not at the Brillouin zone center. 

Such an indirect gap semiconductor that satisfies the criteria for spin
filters in conjunction with some ferromagnetic metals is hexagonal boron
nitride [{\em h}-BN]. It has a layered structure (resembling graphite)
with a hexagonal Bravias lattice. Its in-plane lattice parameter of
2.504{\AA}  is a very good match to the nearest neighbor distances
2.507{\AA}, 2.492{\AA} and 2.506{\AA} in the hexagonal atomic layers of
hcp Co, fcc Ni and fcc Co, respectively. Because of this accurate lattice
matching, monolayers of hexagonal boron nitride grown on (111) surfaces
of fcc Ni are highly ordered and in atomic registry with the
substrate\cite{Nagashima}. Because BN consists of light atoms the effects
of spin-orbit coupling in BN should be very weak, which should be
advantageous in a candidate for an ideal spin filter; see Section
\ref{Conclusions}.
 
For fcc and hcp crystals in {\em exact} atomic registry at the interface
with a hexagonal basal plane of {\em h}-BN, the projections of the
reciprocal lattice vectors of the fcc and hcp crystals onto the interface
coincide with the projections of the {\em h}-BN reciprocal lattice
vectors onto the interface and are indicated by the filled circles in
Fig.\ref{fig02}. The projections of the first Brillouin zones of the hcp
crystal and of {\em h}-BN onto the interface are indicated by the hexagon
in the figure. Recent band structure calculations\cite{Catellani,Xu,Ma}
indicate that {\em h}-BN is an indirect gap semiconductor with the lowest
conduction band minimum at M and the valence band maximum at H or K.
Combining these results with those in the literature
\cite{Papaconst,Mertig,web} for the band structures and Fermi surfaces
of  hcp Co, fcc Ni and fcc Co, and applying the criteria developed in
Section
\ref{Rules} yields the following predictions: 

The interface between a hexagonal atomic layer of hcp Co and a hexagonal
basal plane of {\em h}-BN should be a near ideal spin filter for both
n-type and p-type {\em h}-BN. The interface between a (111) atomic plane
of fcc Ni or fcc Co and a hexagonal basal plane of {\em h}-BN should be a
near ideal spin filter for p-type {\em h}-BN but not for n-type {\em
h}-BN. In each case, minority spin band electrons are predicted to be
transmitted by the filter.

\subsubsection{Boron Nitride with the zinc blende and 
wurtzite crystal structures}

While {\em h}-BN is the stable form of boron nitride under normal
conditions, BN can also exist with the zinc blende and wurtzite crystal
structures, and band structure calculations have been performed for those
systems as well \cite{Xu}. Both are indirect gap materials with the
valence band maximum at $\Gamma$. The conduction band minima are at X for
the zinc blende form and at K for the wurtzite \cite{Xu}. The hexagonal
atomic planes of these materials should lattice-match reasonably well to
the hexagonal atomic planes of hcp Co, fcc Ni and fcc Co. 

The present theory makes the following predictions for these interfaces
based on the Co and Ni band structures in Refs.
\cite{Papaconst,Mertig,web}: The interfaces between the hcp Co and both
the zinc blende and wurtzite BN semiconductors should be near ideal spin
filters for both the p-type and n-type semiconductors. This should also
be true of interfaces between the fcc Ni and Co and the wurtzite form of
BN. However the interfaces between the fcc Ni and Co and the zinc blende
form of BN should be near ideal spin filters for the p-type BN but not for
the n-type BN. In the case of fcc Co with p-type BN having the zinc
blende or  wurtzite structures see also Ref.\cite{exception}.

\subsubsection{Strained Germanium (111) films}

Another indirect gap semiconductor that meets the criteria for a near
ideal spin filter is (n-type or p-type) Ge in the form of a thin, highly
strained film whose (111) face is in atomic registry with a (001) face of
hcp Co or a (111) face of fcc Ni or Co. In this case the tensile in-plane
strain experienced by the Ge would lift the degeneracy between its
conduction band minima   and the lowest conduction band minimum (at L in
the [111] direction) would satisfy  the wave vector selection rule
criteria in Section \ref{Rules}, as would the valence band maximum at 
the zone center\cite{exception}.

\subsection{Spin filters involving the ferromagnet CoS$_2$}

The above examples of spin filters all involve hcp Co, fcc Ni or fcc Co
as the ferromagnetic metal. Another ferromagnetic metal that
will be shown below to be an interesting candidate for spin filters is 
CoS$_2$ which has a
simple cubic Bravais lattice with a lattice parameter
$a=5.407$\AA. The lattice parameters of several semiconductors with the
zinc blende  and diamond crystal structures  have values very close to
this, so that accurate lattice matching at interfaces between (001)
crystal planes of those semiconductors  and a (001) plane of CoS$_2$ is
possible. Fig.\ref{fig03} shows the projections onto a (001) interface
plane of the reciprocal lattice vectors of CoS$_2$ (open and full
circles) and of a semiconductor with the zinc blende  or diamond crystal
structure (open circles) exactly lattice-matched to the CoS$_2$ at the
interface. The square is the projection of the first Brillouin zone of
CoS$_2$. X$_{\rm S}$ and L$_{\rm S}$ denote projections of some X and L
points of the semiconductor Brillouin zone onto the interface plane. X, R
and M denote projections of the respective points of the CoS$_2$
Brillouin zone onto the interface plane.  Band structure calculations
\cite{Zhao} indicate that the projection of the  majority electron Fermi
surface of CoS$_2$ onto the (001) plane occupies most of  the projection
of the first Brillouin zone onto the (001) plane but not the immediate
vicinity of the corners, whereas the projection of the minority Fermi
surface forms diagonal cross-shaped regions centered on the corner points
R. Thus for a semiconductor whose Fermi surface is at the center of the
Brillouin zone or at or near the zone boundary point X, according to the
criteria in Section
\ref{Rules}, only majority spin band electrons are transmitted at the
Fermi energy from CoS$_2$ into the semiconductor through the lattice
matched (001) interface, while for a semiconductor whose Fermi surface is
at or near the zone boundary point L, only minority spin band electrons
are transmitted at the Fermi energy from CoS$_2$ into the semiconductor
through the lattice matched (001) interface. Thus most semiconductors
with the zinc blende or diamond structures whose lattice parameters match
CoS$_2$ at a (001) interface should be considered candidates for spin
filters with CoS$_2$.

Some examples of semiconductors with the zinc blende and diamond
structures that approximately lattice-match CoS$_2$ in the above sense
(and thus are candidates for spin filters with CoS$_2$ and a (001)
interface) are listed below.  The name of the semiconductor is followed
in parentheses by the ratio $a_{\rm p}/a$ of the value $a_{\rm p}$ of the
lattice parameter required for a perfect match to CoS$_2$, to the actual
value
$a$ of the lattice parameter for the semiconductor: CuCl(1.000),
ZnS(0.999), Si(0.996), GaP(0.993), AlP(0.989), AlAs(0.959), Ge(0.956),
GaAs(0.956), ZnSe(0.954), CuBr(0.950), CdS(0.928), InP(0.921).

Note that the technologically important semiconductor Si is a very good
lattice match for CoS$_2$, and that the matching may be improved further
by alloying the Si with a small amount of C or working with thin films
and a slightly strained interface.

\subsection{Spin filters involving FeAl, $\tau$-MnAl, Pd$_3$Fe or 
Co$_3$Pt}

Their calculated band structures\cite{Sundararajan,Prins} indicate that
both FeAl and $\tau$-MnAl do not satisfy equation (\ref{eq:metsem0})
for some combinations of interface orientation and spin, 
a necessary but not  sufficient
condition for a ferromagnetic metal to be an ideal spin filter in
conjunction with a semiconductor whose Fermi surface is close to the
center of the Brillouin zone. FeAl is a cubic material and does 
not satisfy equation
(\ref{eq:metsem0}) for the majority spin Fermi surface and 
(110) interfaces. 
Its lattice parameter is close
to one-half of those of several semiconductors with zinc
 blende or diamond crystal structures so
that a good lattice match between FeAl and those semiconductors 
at a (110) interface is possible.
$\tau$-MnAl has a tetragonal structure. Its lattice parameter 
in the plane with
four-fold symmetry is also close to one half of those of some
semiconductors with zinc blende and diamond crystal structures, so that a
good lattice match with (001) faces of those semiconductors is possible.
The minority spin Fermi surface of $\tau$-MnAl does not satisfy equation 
(\ref{eq:metsem0}) for these interfaces.
But  a detailed analysis of these systems, taking account of 
non-zero reciprocal lattice vectors in
equations (\ref{eq:metsem}) and (\ref{eq:metsemnot0}), 
shows that they do not satisfy the 
criteria for ideal spin filters for
semiconductors whose Fermi surfaces are close to the center of 
the Brillouin zone or to X. The
published Fermi surface data
\cite{Sundararajan,Prins} is not complete enough to decide whether these
systems should be near-ideal spin filters with semiconductors whose Fermi
surfaces are near L. However, for strained thin films of n-type
semiconductors with zinc blende or diamond structures and Fermi surfaces
near X in the Brillouin zone, and in atomic registry with $\tau$-MnAl at
a (001) interface as above, if the interfacial strain is such that it
lowers the conduction band minima at ($\pm k$,0,0) and (0,$\pm k$,0)
relative to those at (0,0,$\pm k$) sufficiently that the conduction band
minima at (0,0,$\pm k$) are emptied of electrons then the (001) interface
is predicted to be an ideal spin filter at low temperatures and bias
voltages. This is because the majority spin Fermi surface of $\tau$-MnAl
intersects the line RX at the edge of the Brillouin zone while the
minority spin Fermi surface does not\cite{Prins}. 

Pd$_3$Fe is a ferromagnetic metal with the Cu$_3$Au crystal structure and
a simple cubic Bravais lattice. The calculated band structure of this
material\cite{Kuhnen} satisfies equation 
(\ref{eq:metsem0}) for both the majority and minority spin Fermi surfaces.
Therefore Pd$_3$Fe
does not satisfy the criteria for ideal spin filters with semiconductors
whose Fermi surfaces are near the center of the Brillouin zone. However,
the size of the lattice parameter of Pd$_3$Fe is close to a factor of
$\sqrt{2}$ smaller than those of some semiconductors with the zinc blende
and diamond crystal structures, making an approximate lattice match at a
(001) interface possible with a
$45^\circ$ relative rotation of the Bravais lattices about the (001)
axis. For strained thin films of such n-type semiconductors with
conduction band minima at X in atomic registry with Pd$_3$Fe, the (001)
interface is again predicted to be an ideal spin filter at low
temperatures and bias voltages provided that the interfacial strain
shifts the conduction band minima in the same way as is described above
for $\tau$-MnAl systems. This is because the  spin down Fermi surface of
Pd$_3$Fe intersects the line RM at the edge of the Brillouin zone while
the spin up Fermi surface does not\cite{Kuhnen}.

According to recent band structure calculations\cite{Kashyap}, 
equation (\ref{eq:metsem0}) is satisfied for both the majority 
and minority spin
Fermi surfaces of Co$_3$Pt, 
the majority spin Fermi surface
intersects the line MX at the edge of the Brillouin 
zone while the minority spin Fermi surface
does not, and the line the line RM intersects both Fermi
surfaces. Based on this, (001) interfaces (similar to those
described above for Pd$_3$Fe) with atomic registry between Co$_3$Pt and
semiconductors with the zinc blende and diamond crystal structures, are
predicted to be ideal spin filters for semiconductors whose Fermi
surfaces are close to L but not for semiconductors whose Fermi surfaces
are close to X or close to the center of the first Brillouin zone.

\subsection{Spin filters involving Gd or Tb}

Gd and Tb are ferromagnetic metals with the hcp structure. Their lattice
parameters in the (001) basal plane with hexagonal symmetry are 3.6336\AA
and 3.6055{\AA} respectively, an  approximate match to the hexagonal
planes of several semiconductors with the zinc blende, diamond and
wurtzite structures. The Fermi surfaces of Gd and Tb are still not
completely understood; the present analysis is based on the Fermi surface
calculations of Ahuja {\em et al.}\cite{Ahuja} These calculations
suggest that for Gd and Tb and (001) interface planes equation 
(\ref{eq:metsem0}) is satisfied neither on the minority spin Fermi surface
nor on the majority spin Fermi surface, and that only the Fermi
surface for the majority spin electrons is present at high symmetry lines
HK, KM and LM on the Brillouin zone boundary. Based on this,
semiconductors lattice matched as above and having Fermi surfaces that
project onto the interface plane near the edges of the projection of the
Brillouin zone of Gd and Tb should be candidates for near ideal spin
filters with Gd and Tb. Some examples are Si(0.946,0.939),
AlP(0.940,0.933), GaP(0.943,0.936), BAs(1.076,1.067) where the numbers in
parentheses indicate the accuracy of the match for Gd and Tb,
respectively. The accuracy of the matching is only fair but may be
improved in strained epitaxial thin films or by semiconductor alloying.

\subsection{Spin filters involving magnetic superlattices}

Ferreira  {\em et al.} have recently suggested that a that a pair of
magnetic superlattices connected by a conducting medium 
with a low carrier density
should function as a perfect spin filter\cite{superlat2} if equation 
(\ref{eq:metsem0}) is satisfied on the majority spin Fermi surface
of the superlattice but not on the minority spin Fermi surface or 
vice versa.  However they
did not identify a material that may serve as their conducting medium
with  a low carrier density.  

In this section I examine the possibility that the interfaces between
some ferromagnetic superlattices and semiconductors may be ideal spin
filters for transmission of spin polarized electrons from the
superlattice to the semiconductor, and also the possibility that
semiconductors may be suitable conducting media with low carrier
densities for devices of the type proposed by Ferreira {\em et al.} In
both cases it turns out to be necessary to go beyond consideration of
equation 
(\ref{eq:metsem0}) and I base the analysis on the theory of Section
\ref{Rules}. 

The calculated minority spin Fermi surface of the (100) oriented superlattice
Fe$_4$/Cr$_4$\cite{Schepsuperl} satisfies equation 
(\ref{eq:metsem0}) for (100) interfaces while the majority spin Fermi surface does not. 
The superlattice has a lattice
parameter in the (100) plane that is very close to one half of those of GaAs, AlAs,
Ge, ZnSe and CuBr, so that very good atomic registry between
Fe$_4$/Cr$_4$ and these semiconductors at a (001) interface is possible.
The analysis of these interfaces as candidates for nearly ideal spin filters in
terms of the criteria of Section \ref{Rules} is as follows: The
projection onto the interface plane of a reciprocal lattice vector of the
semiconductor Bravais lattice connects the projection of the center of
the Brillouin zone of the Fe$_4$/Cr$_4$ superlattice to the projection of
the corner of the Brillouin zone. Assuming that the electronic structure
of the superlattice is as in Ref.\cite{Schepsuperl}, this implies that
the transmission of neither majority nor minority spin electrons is
forbidden from this superlattice to semiconductors (lattice matched to
the superlattice as above) whose Fermi surfaces are near the center of
the Brillouin zone or near X. Furthermore, the projection of neither the
majority nor the minority spin Fermi surface of the superlattice is
connected by the projection of a reciprocal lattice vector to the
projection of a semiconductor Fermi surface located at L. Thus despite
the (100) Fe$_4$/Cr$_4$ superlattice satisfying equation 
(\ref{eq:metsem0}) for one spin species and not for the other, its
calculated electronic structure
\cite{Schepsuperl} indicates that it is not suitable for use in spin
filters with semiconductors lattice matched in this way, except possibly
for strained thin semiconductor films with Fermi surfaces near X, as has
been described above for $\tau$-MnAl. The same applies to the use of such
semiconductors as the conducting medium with  low carrier density in devices
of the type proposed by Ferreira {\em et al.} with the (100)
Fe$_4$/Cr$_4$ superlattice. 

Some other magnetic superlattices whose calculated electronic
structures\cite{Sanchosuperl} satisfy equation 
(\ref{eq:metsem0}) on their spin down Fermi surfaces but not on their spin 
up Fermi surfaces are Ni$_n$Co$_m$ multilayers grown along the (111)
direction. Spin filtering by interfaces between these superlattices and
various semiconductors can be  analyzed using the results of Section
\ref{Rules} (consideration of equation 
(\ref{eq:metsem0}) alone is again
insufficient) in a similar way to the interfaces of those semiconductors
with Ni and Co that are treated in Section
\ref{CoNi}, with similar results. Thus such superlattices should, like Co
and Ni crystals, be suitable for injecting spin polarized electrons into
the same semiconductors. Devices of the kind proposed by Ferreira {\em et
al.} based on these superlattices and semiconductors should also in
principle be possible.

\section{Disorder at the interface}
\label{Disorder}

The preceding sections have addressed ordered interfaces between 
ferromagnetic metals and semiconductors. Here I will consider the effects
of disorder at the interface  on spin filters (focussing particularly on
intermixing disorder) by  solving a simple tight binding model
numerically. For a perfectly ordered interface  and a partly
spin-polarized ferromagnetic metal, the model exhibits a regime in which
the electron transmission through the interface is completely spin
polarized, i.e., the interface is a perfect spin filter. On the other
hand, for strong intermixing between the metal and semiconductor at the
interface, a regime occurs in which the spin polarization of the
electrons transmitted into the semiconductor is much less than the spin
polarization of the electrons at the Fermi level in the ferromagnetic
metal. I.e., mixing disorder at the interface can make the interface a
spin-``anti-filter'' by strongly spin-{\em de}polarizing the electric
current transmitted from the ferromagnet to the semiconductor, even in
the absence of spin flip scattering. This effect effect has a different
origin from that described by Schmidt {\em et al.}\cite{Schmidt} but
should work in concert with the latter. 

\subsection{Model}  

The geometry of the model to be considered is shown in Fig.\ref{fig1}(a):
Semi-infinite crystals of ferromagnet and semiconductor meet at a plane
interface $d$ layers thick where mixing between the ferromagnet and
semiconductor occurs.  It is assumed that in each of these
$d$ layers the semiconductor and ferromagnetic species are randomly
distributed and that the average concentration of each species varies
linearly with position through the interface. The model electronic
Hamiltonian is
\begin{equation}
 H= \sum_{i,\sigma}\epsilon_{i\sigma} a^\dagger_{i\sigma} a_{i\sigma} -
\sum_{i,j,\sigma} t_{ij} a^\dagger_{i\sigma} a_{j\sigma} 
\label{eq:H}
\end{equation}
where $a^\dagger_{i\sigma}$ creates an electron with spin $\sigma$ on
site $i$ of the (simple cubic) lattice and $t_{ij}$ is the  hopping
matrix element between sites $i$ and $j$. The form of site energy
$\epsilon_{i\sigma}$ is illustrated in Fig.\ref{fig1}(b): In the
semiconductor region it is a constant independent of the spin
$\epsilon_{i\sigma} = \epsilon_{s}$. In the ferromagnet it takes values
$\epsilon_{i\sigma} = \epsilon_{m} \pm f/2$ where $f$ is the energy
splitting between minority and majority spin electron bands.  It is
assumed that the interface is an ohmic contact and that in the interface
region where the ferromagnetic and semiconductor species mix the site
energies for the minority and majority spin electrons are given by the
mean field form $\epsilon_{i\sigma} = \epsilon_{s}
\pm \alpha f/2$ or $\epsilon_{i\sigma} = \epsilon_{m} \pm \alpha f/2$ if
the site is occupied by the semiconductor or ferromagnetic species
respectively; 
$\alpha$ is the average concentration of the ferromagnetic species in the
the interface layer in which
$\epsilon_{i\sigma}$ is evaluated. The physical meaning of this
assumption is that the electron-electron interaction effects that give
rise to the energy splitting between the majority and minority spin
electrons have a range of at least a few lattice sites and in the mixing
region affect electrons on sites occupied by the semiconductor species as
well as those occupied by the ferromagnetic species. The hopping matrix
elements are assumed to be nearest neighbor and of the form $t_{ij} =
t_s$ if
$i$ and $j$ are both semiconductor sites, $t_{ij} = t_m$ if $i$ and $j$
are both ferromagnetic sites, and $t_{ij} = (t_s + t_m)/2$ if one of the
sites is ferromagnetic and the other semiconductor.

\subsection{Theoretical Considerations and Method of Solution}  

According to the Landauer theory of transport\cite{Lan57} 
the electrical conductance $G$ of a structure such as that
in Fig.\ref{fig1}(a) is given by $G = e^2 /h
\sum_{kl\sigma} T^\sigma_{kl}$ where
$T^\sigma_{kl}$ is the probability that an electron with spin $\sigma$
incident 
from the source (ferromagnet) in channel $l$ at the Fermi energy is
transmitted into channel $k$ of the drain (semiconductor). 
Thus the spin-dependent Landauer transmission probabilities $T^\sigma =
\sum_{kl} T^\sigma_{kl}$ are the appropriate measure of how well spin
up and down electrons are transmitted through the interface and are
studied in the present work. 
The results for $T^\sigma$ presented here
have been obtained by solving the Lippmann-Schwinger equation
\begin{equation}
 \Psi^\sigma _l = \Phi^\sigma _l + G_0(E+i\epsilon)V\Psi^\sigma _l 
\label{eq:LS}
\end{equation}
where $G_0(z)=(z-H_0)^{-1}$ and $H_0$ are the Green's function and
Hamiltonian for the system shown in Fig.\ref{fig1}(a) but with no
disorder present and with the ferromagnet decoupled from the
semiconductor. I.e.,
$H_0$ is defined similarly to the Hamiltonian $H$ given by equation
(\ref{eq:H}) but with the width $d$ of the intermixing region equal to
zero and with $t_{ij} = 0$ if one of the
sites $i,j$ is in the ferromagnet and the other in the semiconductor.
$V \equiv H-H_0$ then contains the coupling between the ferromagnet
and the semiconductor and any disorder that is present in the system.
In equation (\ref{eq:LS}) $\Phi^\sigma _l$ is the eigenstate of $H_0$
that corresponds to an electron with energy $E$ and spin $\sigma$
that travels from left to right in 
channel $l$ of the (semi-infinite) ferromagnet  and is reflected at the
interface where the coupling to the semiconductor has been switched off
($t_{ij} = 0$) in $H_0$. $\Psi^\sigma _l$ is the corresponding
eigenstate of the complete Hamiltonian $H$ that is partly reflected
at the interface and partly transmitted into the semiconductor. 
$G_0$
and $\Phi^\sigma _l$ were evaluated analytically.
$\Psi^\sigma _j$ was then evaluated by solving equation
(\ref{eq:LS}) numerically using matrix techniques. In
the semiconductor region $\Psi^\sigma _j$ was expressed in terms of
its partial transmission amplitudes
$\tau^\sigma_{kl}$ into the various semiconductor channels $k$. The
partial transmission probabilities $T^\sigma_{kl} = |\tau^\sigma_{kl}|^2
v^\sigma_k/v^\sigma_l$ that enter $T^\sigma$ were obtained using
the calculated propagation velocities $v^\sigma_k$ and $v^\sigma_l$ of
electrons at the Fermi energy with spin $\sigma$ in channels $k$ and $l$
of the semiconductor and ferromagnet, respectively.

\subsection{Results}
 
Representative results are shown in
Fig.\ref{fig2}. Here the metal and semiconductor regions of
Fig.\ref{fig1}(a) are semi-infinite nanowires with a cross-section
of 15$\times$15 lattice sites. $\epsilon_{m}=0$ and $t_m > 0$. The
exchange splitting between the majority and minority spin bands in the
ferromagnet is three times the electron hopping
parameter in the ferromagnet, $f=3t_m$. The semiconductor
conduction band width is half of the band width
of the metal, $t_s = t_m /2$. The Fermi energy $E_F$ is in units of
$t_m$.  It should be noted that the essential qualitative
properties of the results to be presented below are insensitive to the
model parameters such as the cross-section of the wire, the size of the
exchange splitting
$f$ and the relative sizes of the band-width parameters $t_s$ and 
$t_m$ in the ferromagnet and semiconductor and the conclusions
drawn (other than those regarding mesoscopic fluctuations) 
will also apply to 
metal-semiconductor interfaces with areas that are macroscopic in size.    

The overlapping majority and minority
spin bands of the ferromagnet can be seen in Fig.\ref{fig2}(a) which
shows the number $n^\sigma_m$ of majority spin (solid line) and
minority spin (dotted line) Landauer channels in the
ferromagnet at the Fermi energy $E_F$ as a function of $E_F$. 

Fig.\ref{fig2}(b) shows the calculated
Landauer transmission probabilities $T^\sigma$ for majority (solid
line) and minority (dotted) spin electrons from the ferromagnet to
the semiconductor as a function of the Fermi energy, for an perfectly
clean, sharp interface with no intermixing of the metal and
semiconductor ($d=0$ in Fig.\ref{fig1}(a)).  For each value of $E_F$,
the value of the site energy $\epsilon_s$ in the semiconductor has
been chosen so that the electron density in the semiconductor
conduction band (and the number $n^\sigma_s$ of
conducting channels at $E_F$ in the semiconductor for each spin
$\sigma$) is small relative to typical values for the metal; the
results shown are for $n^\sigma_s = 6$.

In region A of Fig.\ref{fig2}(b) ($-7.5 t_m < E_F < -4.5 t_m$) only
majority spin electrons are transmitted into the semiconductor
because only they are present in the ferromagnet. In region B ($-4.5
t_m < E_F < -3.1 t_m$) both majority and minority spin electrons are
transmitted into the semiconductor. In region C ($-3.1 t_m <
E_F < -0.1 t_m$) {\em although both majority and minority spin
electrons are present in the ferromagnet at the Fermi energy, 
only minority spin electrons are transmitted into the semiconductor} :
The majority spin electrons are reflected perfectly at the
metal-semiconductor interface and the system is an ideal spin filter
-- electron injection into
the semiconductor is 100\% minority spin-polarized. In region D ($E_F >
-0.1 t_m$) both the majority and minority spin electrons are reflected
perfectly at the interface and neither species is transmitted into the
semiconductor. 

The 100\% spin-polarization of the electrons transmitted into the
semiconductor in region C is clearly not due to the difference
between the majority and minority spin densities of states in the
ferromagnet since the number of majority spin channels at the Fermi
energy {\em exceeds} (see Fig.\ref{fig2}(a)) that of the minority spin
channels throughout this energy range in which only the minority
spin carriers are transmitted the semiconductor. 
It is due instead to the
selection rule associated with conservation of the component of the
electron wave vector parallel to the interface between the
ferromagnet and semiconductor \cite{rlv} and to the general property of
semiconductors that under near-equilibrium conditions the conduction
band electrons are confined to {\em very small} regions of
$k-$space near the conduction band minima: In this model the electron
eigenenergies in the ferromagnet are $E^\sigma _{\bf k} =
\epsilon_m -2t_m (cos(k_x a) + cos(k_y a)+cos(k_z a)) \pm f/2$
where $a$ is the lattice parameter. The Fermi energy is close to
the semiconductor conduction band minimum which in the present
model is at ${\bf k}= {\bf 0}$. With the interface perpendicular
to the $z$-axis, conservation of the components of the
wave-vector parallel to the interface requires that only
electrons with $k_x$ and $k_y$ close to zero can be transmitted
into the semiconductor at the Fermi energy. For $k_x = k_y =0$,
$E^\sigma _{\bf k} = \epsilon_m -2t_m (2+cos(k_z a)) \pm f/2$
which implies that $E^\sigma _{\bf k} \le \epsilon_m -2t_m \pm
f/2$. This means that only electrons with energies less than
$\epsilon_m -2t_m \pm f/2$ can be transmitted from the
ferromagnet to states at the bottom of the semiconductor
conduction band. For the model parameters $\epsilon_m=0$ and
$f=3t_m$ chosen in Fig.\ref{fig2}, this implies that majority and
minority spin electrons can be transmitted from the ferromagnet
to states at the bottom of the semiconductor conduction band at
energies below $-3.5t_m$ and $-0.5t_m$, respectively. The
corresponding high energy cutoffs for transmission of majority and
minority carriers in Fig.\ref{fig2}(b) are slightly higher, at   
$-3.1 t_m$ and $-0.1 t_m$ respectively, because in the
numerical calculations the Fermi energy was chosen
slightly above the semiconductor's conduction band minimum instead of
right at the minimum as in the above analysis.

The calculated transmission probabilities of majority (solid line) and
minority (dotted line) spin electrons from the ferromagnet to the
semiconductor through a disordered interface are shown in
Fig.\ref{fig2}(c). The model parameters are the same as in
Fig.\ref{fig2}(b) except that now the thickness of the interface where
mixing of the semiconductor and metal occurs is $d=8$ lattice layers.
Since the physics of electron transmission through the perfect
interface (Fig.\ref{fig2}(b)) is controlled by a
selection rule associated with lattice periodicity parallel to the
interface, one should expect the strongly disordered interface to
behave differently; the differences between Fig.\ref{fig2}(b)
and Fig.\ref{fig2}(c) are indeed striking: Whereas for the perfect
interface transmission from the ferromagnet to the semiconductor is
partly spin polarized in region B and completely spin polarized in
region C of Fig.\ref{fig2}(b), the transmission is close to being
completely spin un-polarized in the corresponding energy ranges for
the disordered interface, as can be seen in Fig.\ref{fig2}(c).
Here the interface acts as a spin ``anti-filter'' with the
spin polarization of the transmitted current being much less than
even that of the electronic channels 
incident on the interface from ferromagnet. The
differences between the total Landauer transmission 
probabilities of the
majority and minority spin electrons through the disordered interface
in this regime are governed by (pseudo-random) fluctuations of the
transmission probabilities. Such
quantum conductance fluctuations with an amplitude of order $e^2 /h$
that occur as the Fermi energy is varied
are well known in other systems\cite{UCF} 
and are the mesoscopic ``finger-print" of the
specific microscopic configuration of the atoms in the disordered
region. In energy region D these fluctuations are replaced by weak
isolated quantum transmission resonances and whether majority or
minority spin electrons or both are transmitted from the semiconductor
into the ferromagnet at a particular energy is controlled by
the microscopic details of the disorder in the interface.

Thus the intermixing at the interface between a ferromagnetic  metal and
semiconductor may act to strongly suppress  the spin polarization of the
transmitted current. In the present model, this suppression is found to be
insensitive to the choice of the model parameters: Even if there are only
$d=2$ disordered layers present at the interface, the spin polarization
of the electron transmission in region C of Fig.\ref{fig2} is reduced to
$\sim 30\%$. Based on the  considerations of Schmidt {\em et
al.}\cite{Schmidt} it may be difficult to detect a spin polarization of
the electric current in this case in spin-valve resistance measurements
on devices in which the transport through the semiconductor is diffusive.
\section{Summary and Some Further Considerations}
\label{Conclusions}
 
In this article it has been pointed out that certain atomically ordered
interfaces between some ferromagnetic metals and semiconductors should
act as ideal spin filters that transmit only electrons belonging to the
majority spin bands of the ferromagnet or only electrons belonging to the
minority spin bands at the Fermi energy. Criteria  for determining which
combinations of ferromagnetic metal, semiconductor and interface should
have this property have been formulated, and examples of systems that
meet these criteria to a high degree of precision have been described. 

The criteria depend only on the bulk band structures of the semiconductor
and ferromagnetic metal and on the translational symmetries of the
semiconductor, metal and interface. Thus  they do not depend on whether a
Schottky barrier is present at the interface or on the strength of this
barrier. If there is a strong Schottky barrier, then although the the
interface may obey the criteria and be an ideal spin filter at low and
moderate bias, the current that it transmits will be weak. The size of
the Schottky barrier depends on the materials involved, and it is
reasonable to expect that among the many systems that should be nearly
ideal spin filters some will have low Schottky barriers. 

Estimates
of the Schottky barrier heights for some of the systems of interest 
may be obtained using the model of Tersoff\cite{Tersoff} 
which expresses the Schottky
barrier height in terms of semiconductor band gap parameters and a
phenomenological fitting term $\delta_m$ that depends only on the 
metal $m$. A reasonable value for this parameter for Co and Ni is 
$\delta_{Co} = \delta_{Ni} = -0.2$eV which yields Schottky barrier
height estimates of 1.2, 0.6, 0.0, -0.3 and 1.2 eV for interfaces
between n-type ZnTe, GaSb, InAs, CdSe and CuI, respectively,
and Ni or Co. These estimates 
suggest that the spin-filter interfaces between
n-type  InAs or CdSe and Ni or Co may be ohmic for the transmitted
spin species. However, their
reliability is uncertain. For example, $\delta_{Au}$ has been 
estimated to be also -0.2eV\cite{Tersoff}. Based on this, within the 
Tersoff model, the Schottky barrier heights for Ni and Co should
be close to those for Au, but the Schottky barrier height
for Au on n-type CdSe has been measured to be 0.49eV\cite{Sze}.  
 
A method 
commonly used to reduce Schottky barriers is to interdiffuse the metal
and semiconductor. This however breaks the translational symmetries of
the interface and should therefore degrade its spin filtering property.
As has been shown in  Section \ref{Disorder}, such intermixed interfaces
can have spin anti-filtering properties, with the transmitted electrons
being much less spin polarized than even those in the ferromagnetic metal
at the Fermi energy. This, in concert with the mechanism of Schmidt {\em
et al.}\cite{Schmidt}, may help to account for some of the setbacks that
have been encountered in experimental attempts to  inject strongly spin
polarized electrons from ferromagnetic metals into semiconductors. A
potentially better way to reduce Schottky barriers in candidates for
ideal spin filters is to modify the chemistry of the interface by
introducing a suitable adsorbate between  the metal and semiconductor
during growth. Ohmic contacts between Al and InGaAs(001)  have been made
in this way by introducing a Si bilayer between the Al and
InGaAs\cite{ohmic}. If the adsorbate is atomically ordered and its
presence does not change the translational symmetries of the system
parallel to the plane of the interface, then the system with the
adsorbate will still conform to the criteria for an ideal spin filter if
the system without the adsorbate does. Thus this is a promising method
for manipulating Schottky barriers while preserving the spin filtering
property of the interface.
Inserting a suitable atomically ordered intermediate layer between the
semiconductor and ferromagnetic metal may also help to enhance the
degree of atomic order at the interface while preserving its spin filtering
property: Since an ordered commensurate monolayer of {\em h}-BN
is known to grow well on (111) surfaces of fcc Ni\cite{Nagashima}, 
introducing one or
more monolayers of {\em h}-BN at the interface between the semiconductors 
and metals in Section \ref{CoNi} is an interesting possibility 
in this regard.  
A more standard method for suppressing
Schottky barriers is to heavily dope the semiconductor. Since the
concentration of dopant atoms in heavily doped semiconductors is still
much smaller than that of the intrinsic semiconductor species, it is
reasonable to expect the spin filtering property of an interface not to
be degraded greatly by this method of Schottky barrier suppression,
making it  a better choice than interdiffusion of the ferromagnetic metal
and semiconductor. 

While the criteria for ideal spin filters guarantee that only electrons
from the majority spin bands of the ferromagnet or only those from the
minority spin bands are transmitted into the semiconductor at the Fermi
energy, the degree of spin polarization of the electrons  injected into
the semiconductor can be influenced by spin-flip scattering if that
occurs at the interface. Spin-orbit coupling can also result in the
electron states of the majority and minority spin bands of the
ferromagnet being incompletely spin polarized. The nominally spin up
electron eigenstates of semiconductor may also contain an  admixture of
spin down (and vice versa), due to spin orbit coupling. Since the linear
combinations of spin up and spin down in the eigenstates of the
ferromagnet and semiconductor need not in general match, spin-orbit
coupling can limit the degree of spin polarization of the
carriers injected into the semiconductor. Thus it is desirable to choose
materials in which spin-orbit coupling of the relevant states  is
minimal, either because of the low atomic numbers of the constituent
elements (as in the semiconductors BN and Si) or because of the
material's band structure.    

This work was supported by NSERC and by the Canadian Institute for
Advanced Research. A patent application based on this work has been
commenced by Simon Fraser University.

\end{multicols}
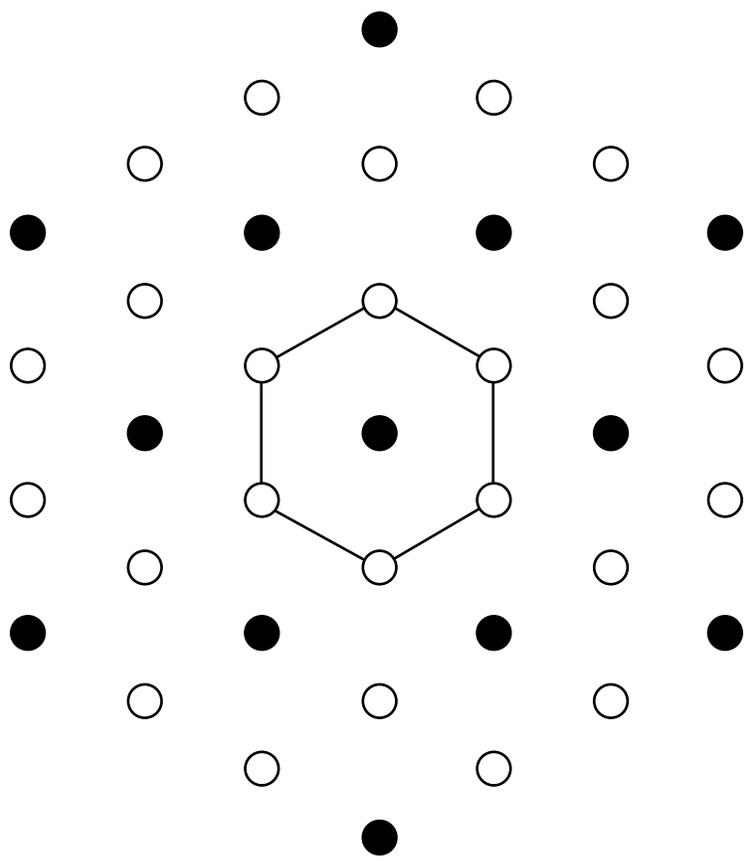
\begin{figure}
\caption{
Schematic of reciprocal lattice vectors and projections of reciprocal
lattice vectors onto the plane of the interface that enter equation
(\ref{eq:metsem}) for interfaces between hcp Co, fcc Ni and fcc Co and
semiconductors with the diamond, zinc blende and wurtzite structures that
are perfectly matched to these metals as described in the text.
Projections of the reciprocal lattice vectors of the metal onto the
interface plane are  represented by filled circles. The projections of
the reciprocal lattice vectors of the semiconductor are indicated by both
open and the filled circles. The open and filled circles also indicate the
reciprocal lattice vectors that are  associated with the group of
symmetry translations (parallel to the interface plane) of the whole
system consisting of the two crystals and the interface between them. The
hexagon is the projection of the boundary of the first Brillouin  zone of
hcp Co onto the interface plane.}
\label{fig01}
\end{figure}
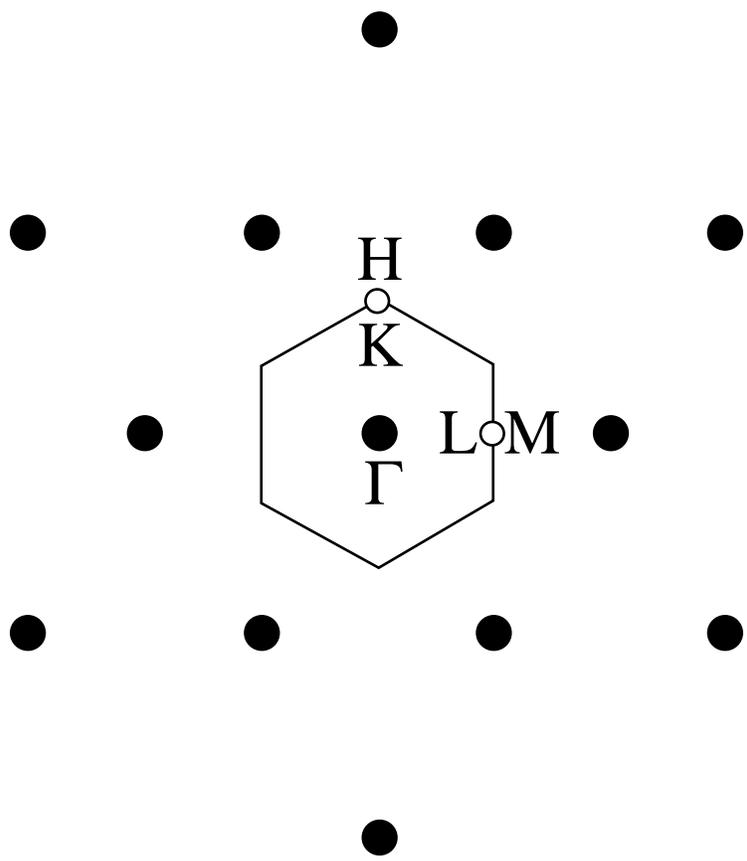
\begin{figure}
\caption{Filled circles represent projections onto a hexagonal {\em h}-BN
basal plane of the reciprocal lattice vectors of hcp and fcc Bravais
lattices perfectly lattice matched to the {\em h}-BN basal plane and also
the projections of the reciprocal lattice  vectors of the {\em h}-BN onto
that plane. The  hexagon is the projections of the first Brillouin zone
of the hcp lattice and of {\em h}-BN onto the same plane; Open circles
indicate the projections of the H, K, L and M points on the Brillouin
zone boundary onto the plane.} 
\label{fig02}
\end{figure}

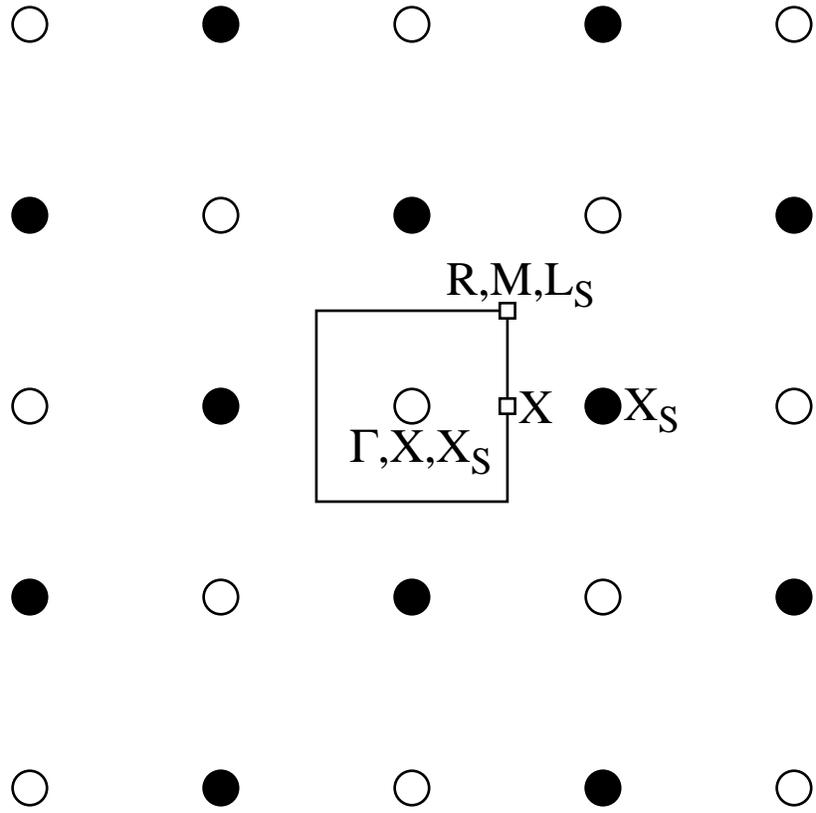
\begin{figure}
\caption{Open circles represent projections onto the interface plane of
the reciprocal lattice vectors of a semiconductor with the zinc blende or
diamond structure perfectly lattice matched at a (001) interface to
CoS$_2$. Open and filled circles represent projections  of the reciprocal
lattice vectors of CoS$_2$ onto the interface plane. The  square is the
projection of the first Brillouin of CoS$_2$ onto the same plane. X$_{\rm
S}$ and L$_{\rm S}$ denote projections of some X and L points of the
semiconductor Brillouin zone onto the interface plane. X, R and M denote
projections of the respective points of the CoS$_2$ Brillouin zone onto
the interface plane.} 
\label{fig03}
\end{figure}

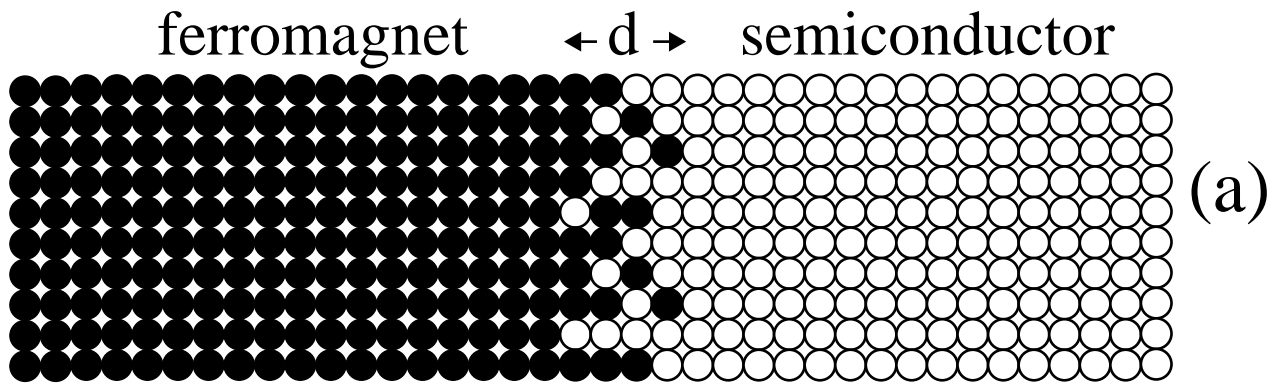
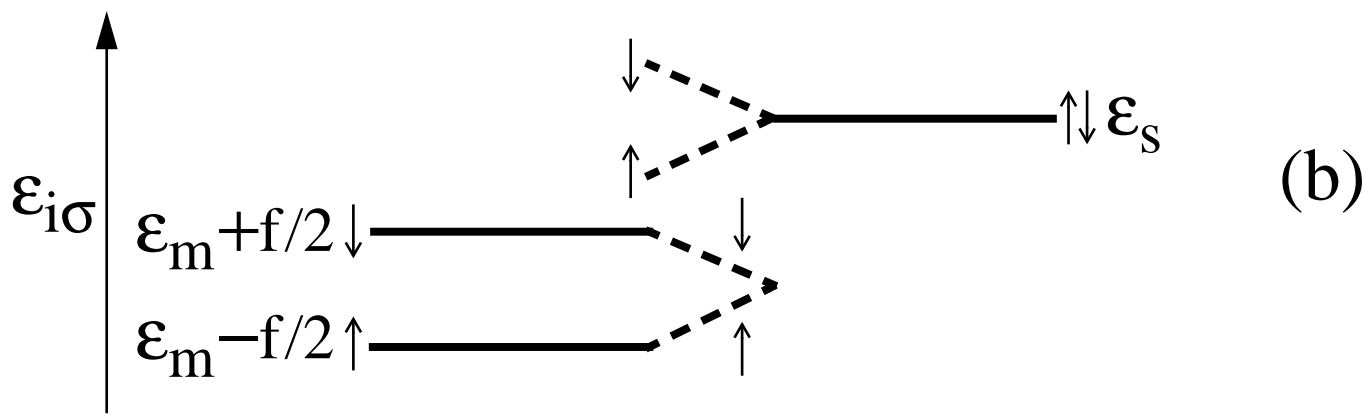
\begin{figure}
\caption{ (a) Schematic of semi-infinite ferromagnetic metal and
semiconductor quantum wires that join at a disordered ohmic interface
$d$ layers thick. (b) Tight-binding model site energies
$\epsilon_{i\sigma}$ for majority
$\uparrow$ and minority $\downarrow$ spin electrons in the ferromagnet
and semiconductor (solid lines) and in the mixed region (dashed lines)
that are shown above in (a).}
\label{fig1}
\end{figure}
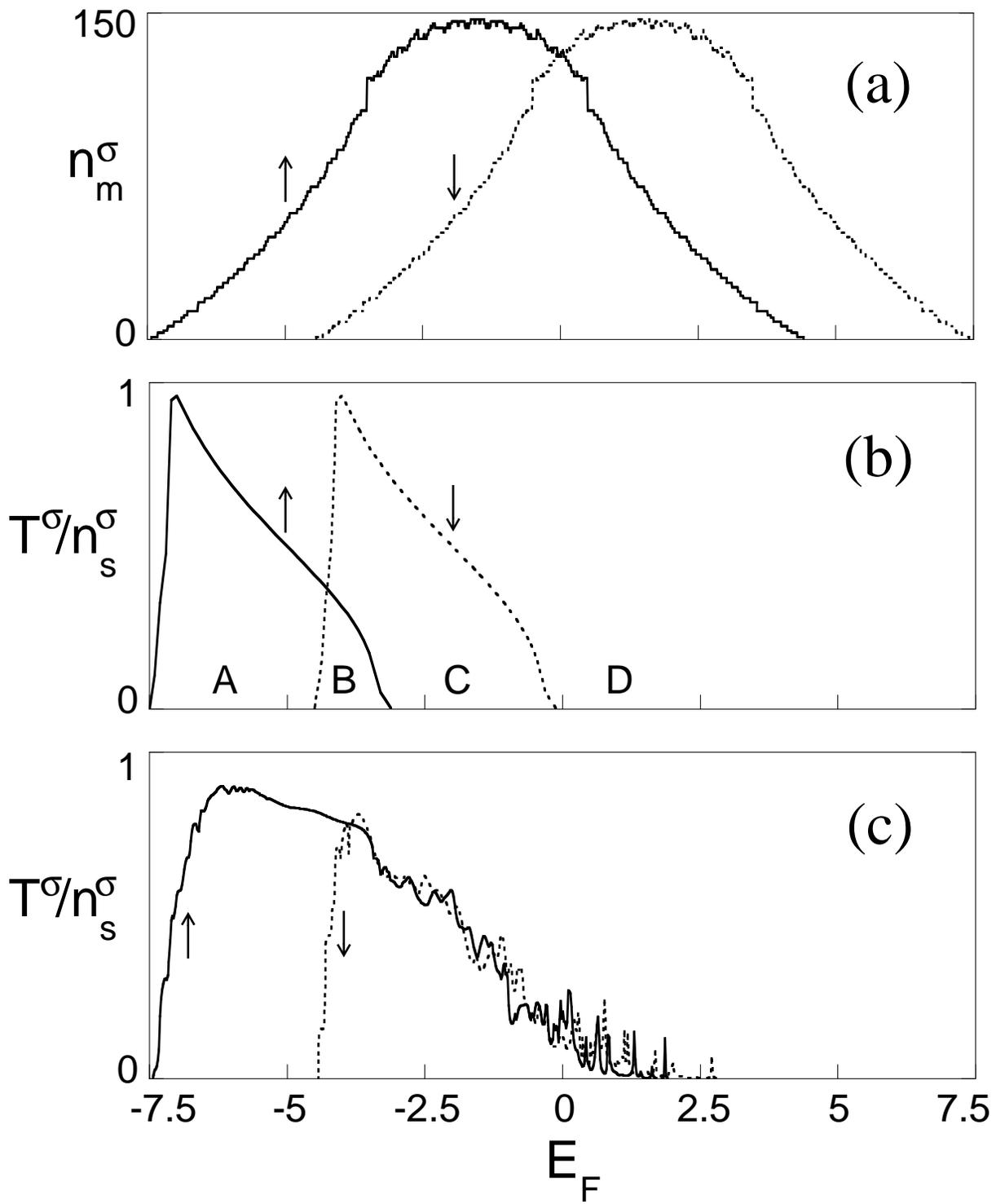
\begin{figure}
\caption{Number of Landauer channels $n^\sigma_m$ in the
ferromagnetic quantum wire at the Fermi energy $E_F$ (a), and
calculated transmission probability $T^\sigma$ from the ferromagnet to
the semiconductor through a perfect (b) and disordered (c) interface,
for majority (solid lines) and minority (dotted) spin electrons at
$E_F$ vs. $E_F$
 for the infinite quantum wire shown in Fig.\ref{fig1}(a).}
\label{fig2}
\end{figure}



\begin{references}
\bibitem{p-type} n-type semiconductors are often regarded
as better candidates for spin injection devices than most p-type
semiconductors since in most semiconductors the valence band is
strongly affected by spin-orbit coupling.
\bibitem{rlv}The reciprocal lattice vectors 
that enter equation (\ref{eq:metsem}) can be set equal to zero
here since the Bravais lattices of the semiconductor and metal in this
model are identical.
\end{references}
\end{document}